\documentclass[preprint,prl,aps,showkeys,showpacs]{revtex4}
\begin{document}
\input{epsf.tex}
\epsfverbosetrue
\title{Bifurcations from stationary to pulsating solitons
in the cubic-quintic complex Ginzburg-Landau equation}
\author{Eduard N. Tsoy\footnote{Also at: Physical-Technical
Institute of the Uzbek Academy of Sciences,
  Tashkent, Uzbekistan} and Nail Akhmediev}
\address{ Optical Sciences Group, Research School of Physical
Sciences and Engineering, \\
The Australian National University, Canberra, ACT 0200, Australia}
\begin{abstract}
Stationary to pulsating soliton bifurcation analysis
of the complex Ginzburg-Landau equation (CGLE) is presented.
The analysis is based
on a reduction from an infinite-dimensional dynamical dissipative system
to a finite-dimensional model. Stationary solitons, with constant amplitude
and  width, are associated with fixed points in the model. For the first time,
{\em pulsating} solitons are shown to be stable {\em limit cycles} in the
finite-dimensional dynamical system. The boundaries between the two types
of solutions are obtained approximately from the reduced model.
These boundaries are reasonably close to those predicted by direct numerical
simulations of the CGLE.
\end{abstract}
\maketitle
 The complex Ginzburg-Landau equation (CGLE) is one of the basic
equations for modelling modulated amplitude waves \cite{Brusch2001},
spatio-temporal dynamics and spontaneous development of coherent
structures in a variety of nonlinear dissipative
systems~\cite{Reviews,Books}. Examples include pulse generation
by passively mode-locked soliton lasers \cite{Ippen94},
signal transmission in all-optical communication lines \cite{bakonyi_2},
travelling waves in binary fluid mixtures \cite{Kolodner},
and also pattern formation in many other physical systems
\cite{Gross}. Complicated patterns  consist of simpler localized
solutions like fronts, pulses, sources and sinks \cite{saarlos92}.

Pulsating soliton solutions of dissipative systems have attracted a
great deal of attention in recent years. They have been found
numerically \cite{Deiss94,AST01,SGGA04} and observed experimentally
\cite{SGGA04} in a fiber laser. Pulsating solitons form one set of
possible localized solutions of the CGLE, and they exist on an equal
basis with stationary solitons.  Such localized waves exist, in
various forms in biology, chemistry and physics.

A pulsating soliton can be be described as a limit cycle of an
infinite-dimensional dissipative dynamical system \cite{Springerbook}.
It is different from the higher-order solitons that are usually
connected with an integrable model \cite{Satsuma74}.
Although numerical simulations show clearly the existence of
pulsating solutions and their bifurcations from stationary solitons,
so far there has been no progress in finding analytic expressions for
pulsating solutions and bifurcation boundaries. The problem is not
simple as there are several parameters of the CGLE that define
the regions of existence for both stationary and pulsating solitons.
Hence, the bifurcation boundaries are surfaces in this multi-dimensional
space of the parameters.

In this work we use a reduction from an infinite-dimensional to a
five-dimensional model, and we aim to find localized solutions of the
CGLE and the transformations that they are subjected to when the
system parameters are varied. Although exact solutions of the CGLE do
exist \cite{Books}, they can be presented explicitly only for certain
relations between the parameters of the equation. Furthermore, only
stationary solutions can be found. Hence, we are faced with the
necessity of finding an efficient approximation to tackle the problem. We have
found that the method of moments, originally developed by Maimistov
\cite{Maim93} for the perturbed nonlinear Schr\"{o}dinger equation
(NLSE) can be used for solving our problem. The moments are the integral
characteristics of the field under consideration. In principle, there
are an infinite number of equations for moments. One can obtain exact results
by using the complete set of these equations. However, in practice, one
uses a trial function with a finite number of parameters, and this is the
way to obtain a significant reduction in the number of variables used for the
description of the dynamics.

The cubic-quintic complex Ginzburg-Landau equation, in dimensionless
form, is written as
\begin{eqnarray}
  i \psi_t + {D \over 2} \psi_{xx} + |\psi|^2 \psi =
  -\nu |\psi|^4 \psi +
\nonumber \\
 i \delta \psi + i \epsilon |\psi|^2 \psi +
  i \beta \psi_{xx} + i \mu |\psi|^4 \psi \equiv R[\psi] \ ,
\label{CGLE}
\end{eqnarray}
where $\psi(x,t)$ is the normalized envelope of the field, $t$ and $x$
are time and spatial variables, respectively, $D$ is the group
velocity dispersion coefficient, $\nu$ is the parameter of the quintic
nonlinearity, $\delta$ represents the linear loss, $\epsilon$ is the
nonlinear gain coefficient, $\beta$ stands for the spectral filtering,
and $\mu$ characterizes the saturation of the nonlinear gain. Stable soliton
solutions of the CGLE exist only for the following choices for the signs of
the coefficients: $\delta, \mu < 0$, $\beta, \epsilon > 0$, and any
sign for $\nu$ and $D$ (see e.g.~\cite{Books,AST01}).  Hence, in this work,
we limit ourselves only to this range.

The method of moments  \cite{Maim93} is a reduction of the complete
problem of the evolution of a field that has an infinite number of degrees
of freedom to the evolution of a finite set of pulse
characteristics. For a localized solution with a single maximum, these
include the peak amplitude, pulse width and center-of-mass position.
For an arbitrary localized field, the two integrals, namely
the energy $Q$ and momentum $P$
\begin{eqnarray}
  Q= \int_{-\infty}^{\infty}{ |\psi|^2 dx}\ , \quad
  P= {1 \over 2} \int_{-\infty}^{\infty}
    { (\psi \psi_{x}^{\ast} - \psi^{\ast} \psi_{x}) dx}.
\label{EM}
\end{eqnarray}
are two basic variables evolving with $t$.
Three higher-order generalized moments, related to the pulse, are given
by the following expressions \cite{Maim93}:
\begin{eqnarray}
  I_1 &=& \int_{-\infty}^{\infty}{x |\psi|^2 dx}\ , \quad
  I_2 = \int_{-\infty}^{\infty}{(x-x_0)^2 |\psi|^2 dx}\ ,
\nonumber \\
  I_3 &=& \int_{-\infty}^{\infty}{(x-x_0)
   (\psi^{\ast} \psi_{x} - \psi \psi_{x}^{\ast}) dx}\ ,
\label{Integr}
\end{eqnarray}
where $x_0(t)= I_1/Q$.
Using Eq.~(\ref{CGLE}) one can derive the evolution equations
for the generalized moments. For the five integrals given above, one
can obtain the following \cite{Maim93}:
\begin{eqnarray}
  && {d Q \over d t} = i \int_{-\infty}^{\infty}
  {(\psi R^{\ast} - \psi^{\ast} R) dx}\ ,
\nonumber \\
  && {d P \over d t} = -i \int_{-\infty}^{\infty}
  {(\psi_x R^{\ast} + \psi_x^{\ast} R) dx}\ ,
\nonumber \\
  && {d I_1 \over d t} = i D P + i \int_{-\infty}^{\infty}
  {x (\psi R^{\ast} - \psi^{\ast} R) dx}\ ,
\label{ODEs} \\
  && {d I_2 \over d t} = -i D I_3 + i \int_{-\infty}^{\infty}
  {(x-x_0)^2 (\psi R^{\ast} - \psi^{\ast} R) dx}\ ,
\nonumber \\
  && {d I_3 \over d t} = 2 P {d x_0 \over d t} +
  i \int_{-\infty}^{\infty} { (2 D |\psi_x|^2 - |\psi|^4) dx } +
\nonumber \\
  && 2 i \int_{-\infty}^{\infty} { (x-x_0)
  (\psi_x R^{\ast} + \psi_x^{\ast} R)} +
  i \int_{-\infty}^{\infty}
  {(\psi R^{\ast} + \psi^{\ast} R) dx}\ .
\nonumber
\end{eqnarray}
Equations~(\ref{ODEs}) are general in the sense that they are
valid for a large class of NLSE-type evolution equations, including
Eq.~(\ref{CGLE}) with arbitrary coefficients as a particular case.
If we use an exact solution of Eq.~(\ref{CGLE}), $\psi$, then
equations~(\ref{ODEs}) are exact.

For problems of a certain class, even the first two Eqs.~(\ref{ODEs})
may be sufficient when one deals with exact two-parameter reductions of the
CGLE solutions \cite{JOSA98}. In Ref.~\cite{JOSA98}, such an approach
was used to find the CGLE solutions in the form of stable soliton pairs
and trains. The method of moments has also been applied to find
stationary solutions of the CGLE~(\ref{CGLE}) in Ref.~\cite{Ostrov04}
although there was no attempt at finding pulsating solitons. Our
tests showed that five is the minimum number of moments
needed to describe pulsating solitons using the reduced system.
Having more of them may improve the accuracy, but the complexity of
the analysis then increases dramatically.

The choice of the trial function is crucial for obtaining solutions with
the desired properties. Any reduction from an infinite-dimensional to
a finite-dimensional system will have deficiencies. In approaches like
this, the choice of the trial function can only be justified at the
last stage of analysis, when the approximate solutions are compared with
numerical simulations of the original equation. We use the fact that
soliton solutions remain localized even when they are pulsating.
Therefore, we take the sech-function:
\begin{equation}
  \psi(x,t)= A\; \mbox{sech} \left( {x- x_0 \over w} \right)
  e^{i[b (x-x_0) + c (x-x_0)^2] }\ ,
\label{Ansatz}
\end{equation}
where $A(t)$, $w(t)$ and $x_0(t)$ are the amplitude, width and maximum
position of the pulse, respectively,
$b(t)$ is the soliton velocity and $c(t)$ is the chirp
parameter. The chirp is highly important, as the numerical
simulations \cite{AST01,SGGA04} show. The number of parameters
in the trial function must correspond to the number of moments
used in the set of equations (\ref{ODEs}). More complicated  trial functions
need more equations in (\ref{ODEs}).

Now, the generalized moments can be expressed in terms of
the variable parameters of the trial function. Evaluation of the
integrals~(\ref{EM}) and (\ref{Integr}), with a help of
Eq.~(\ref{Ansatz}), gives the following expressions:
\begin{eqnarray}
  Q= 2 A^2 w, \quad P= -2 i A^2 w b, \quad  I_1= 2 A^2 w x_0,
\nonumber \\
  I_2= (\pi^2/6) A^2 w^3, \quad I_3= i (2 \pi^2/3) A^2 w^3 c.
\label{Values}
\end{eqnarray}
Then, using Eqs.~(\ref{ODEs}), one can obtain a set of ordinary
differential equations for the soliton parameters
$Q,w,c,x_{0}$ and $b$:
\begin{eqnarray}
  Q_t &=& F_1   \equiv   2 Q \nonumber \\
  & & \times  \left[ \delta - \beta b^2 + {\epsilon \over 3}
  {Q \over w} + {2 \mu Q^2 - 5 \beta \over 15 w^2}
  - {\pi^2 \over 3} \beta c^2 w^2 \right] ,
\nonumber \\
  w_t &=& F_2   \equiv  -{2 \epsilon \over \pi^2} Q + {8 \beta-\mu Q^2 \over
  \pi^2 w} + 2 D c w -
  {16 \pi^2 \over 15} \beta c^2 w^3   ,
\nonumber \\
c_t &=& F_3 \equiv -2 D c^2 - {1 \over \pi^2 w^2}  \label{Model} \\
 & &  \times \left[
  4 \left( {\pi^2 \over 3} + 1 \right) \beta c +
  {Q \over w}
  + {8 \nu Q^2 -30 D \over 15 w^2}  \right] ,  \nonumber\\
\nonumber \\
  x_{0,t} &=& F_4 \equiv   b (D - {2 \pi^2 \over 3}  \beta c w^2) ,
\nonumber \\
  b_t &=& F_5 \equiv  - {4 \over 3} \beta ( {1 \over w^2} + \pi^2 c^2 w^2 )\,
  b  \ .
\nonumber
\end{eqnarray}

Fixed points (FPs) of the dynamical system~(\ref{Model}) can be found from the
set of algebraic equations $F_j =0,\ j=1, \dots,5$.  The stability of
the FPs is determined from the analysis of eigenvalues $\lambda_j$,
$j=1, \dots,5$, of the Jacobian matrix $ M_{ij} = \partial F_i /
\partial p_j $, where $\{p_1, \dots, p_5 \} \equiv \{ Q, w, c, x_0, b
\}$, and $i= 1, \dots, 5$. If the real part of at least one eigenvalue
is positive, then the corresponding fixed point is unstable. In
principle, the whole five-dimensional dynamical system (\ref{Model})
can be studied using the specialized software \cite{AUTO2000}. We analyse
the system based on general theory and using further simplifications.

Firstly, we consider solutions with $b=0$ in Eqs. (\ref{Model}).
The real part of the eigenvalue which corresponds to $b$ is negative for
any set of the system parameters. The soliton center, $x_0(t)$, takes
a constant value for $t \to \infty$ and the real part of the
corresponding eigenvalue is zero (neutral stability). Therefore, we
can make a further reduction and consider a system with only three
variables, viz. $Q, w$, and $c$. We denote the three
corresponding eigenvalues by $\lambda_1$, $\lambda_2$ and
$\lambda_3$. Note that, since the characteristic equation for these
eigenvalues is cubic, then either $\lambda_1= \lambda_2^{\ast}$ and
$\lambda_3$ is real, or all three $\lambda_i$ are real.

\begin{figure}[c]
\centering
\epsfxsize=8.5cm \mbox{\epsffile{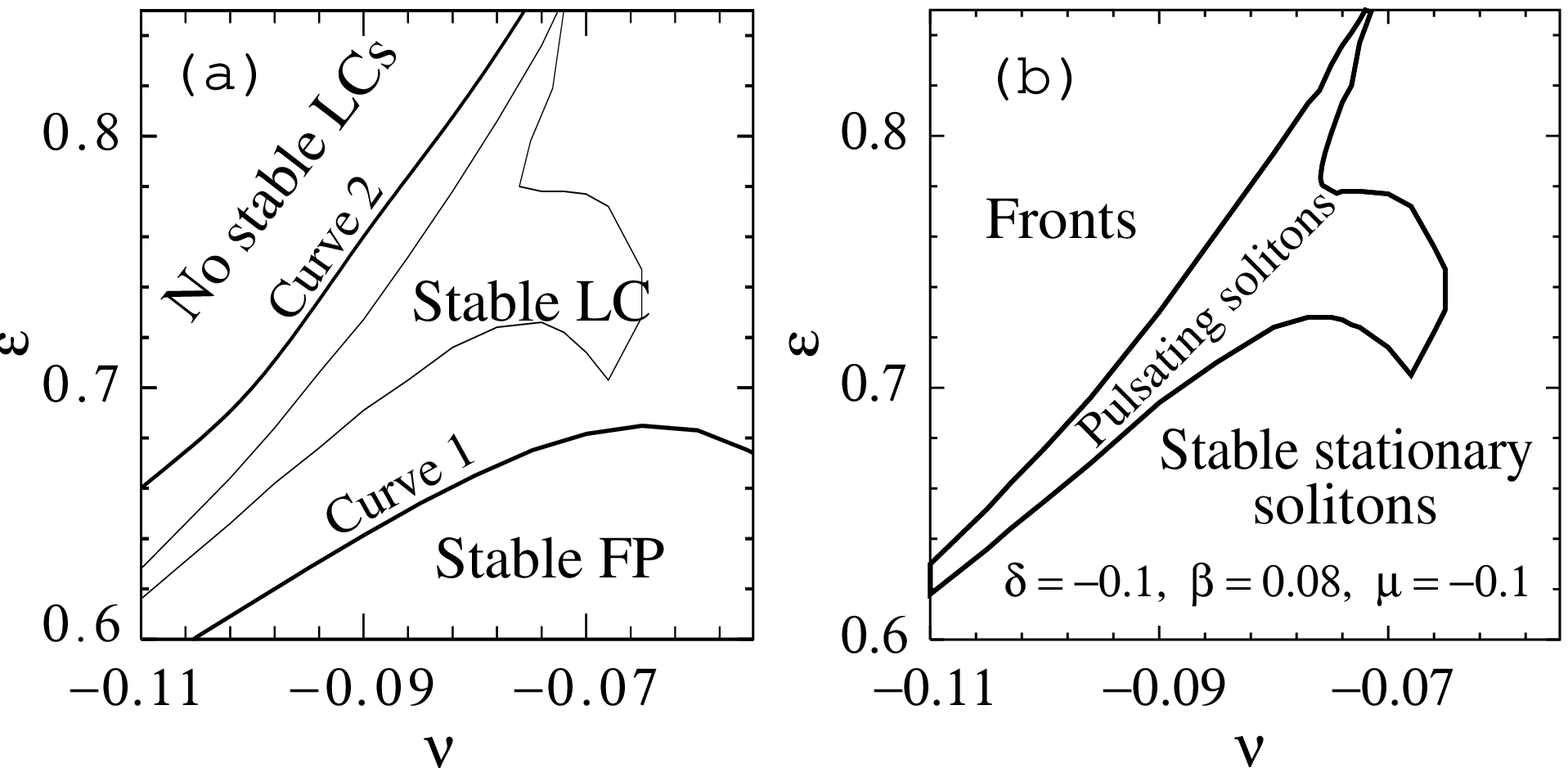}}
\caption{(a)
Regions of existence and stability of FPs and limit cycles (LCs) of the
reduced system in ($\nu, \epsilon$)-plane. The central region, between
the two solid lines $1$ and $2$, corresponds to stable LC. (b) Regions
of existence for various soliton solutions obtained from
 numerical simulations of CGLE (\ref{CGLE}).
The region for pulsating solitons, found numerically in Ref.~\protect
\cite{AST01}, is copied from (b) to (a) for comparison.
The parameters of the dynamical system are shown in (b).} \label{Fig:ne1}
\end{figure}

We find FPs numerically by solving the algebraic equations, and
calculate the characteristic eigenvalues by analyzing the Jacobian
matrix. In this
way, we have identified the regions where solutions have distinctive
features. The bifurcation diagram for the dynamical system ~(\ref{Model}) in
the ($\nu, \epsilon$)-plane is presented in Fig.~\ref{Fig:ne1}.
We take $D=1$ and $\delta= -0.1$ henceforth in the paper.
To facilitate comparison with exact results, the parameters of the model
are chosen to be the same as in the numerical simulations of
Ref.~\cite{AST01}, namely $\mu= - 0.1$ and $\beta=0.08$.

When the value of the gain $\epsilon$ is small, there are no stable or
unstable FPs (i. e. no stationary solitons) in the system. This
region is located below $\epsilon=0.6$, and therefore it is not shown in
Fig.~\ref{Fig:ne1}. For moderate values of $\epsilon$,
there are two FPs. In the region below the curve $1$ in Fig.\ref{Fig:ne1}a,
one FP is stable while the other one is unstable.
The signs of the real and imaginary parts of the eigenvalues,
for the stable FP in this region are the following:
$\mbox{Re}[\lambda_{1,2}] <0$,
$\mbox{Im}[\lambda_{1}] = -\mbox{Im}[\lambda_{2}]$,
$\mbox{Re}[\lambda_{3}] <0$, and $\mbox{Im}[\lambda_{3}] =0$.
The stable FP corresponds to a stationary CGLE soliton with
constant soliton parameters $p_j$.

\begin{figure}[c]
\centering
\epsfxsize=8.5cm \mbox{\epsffile{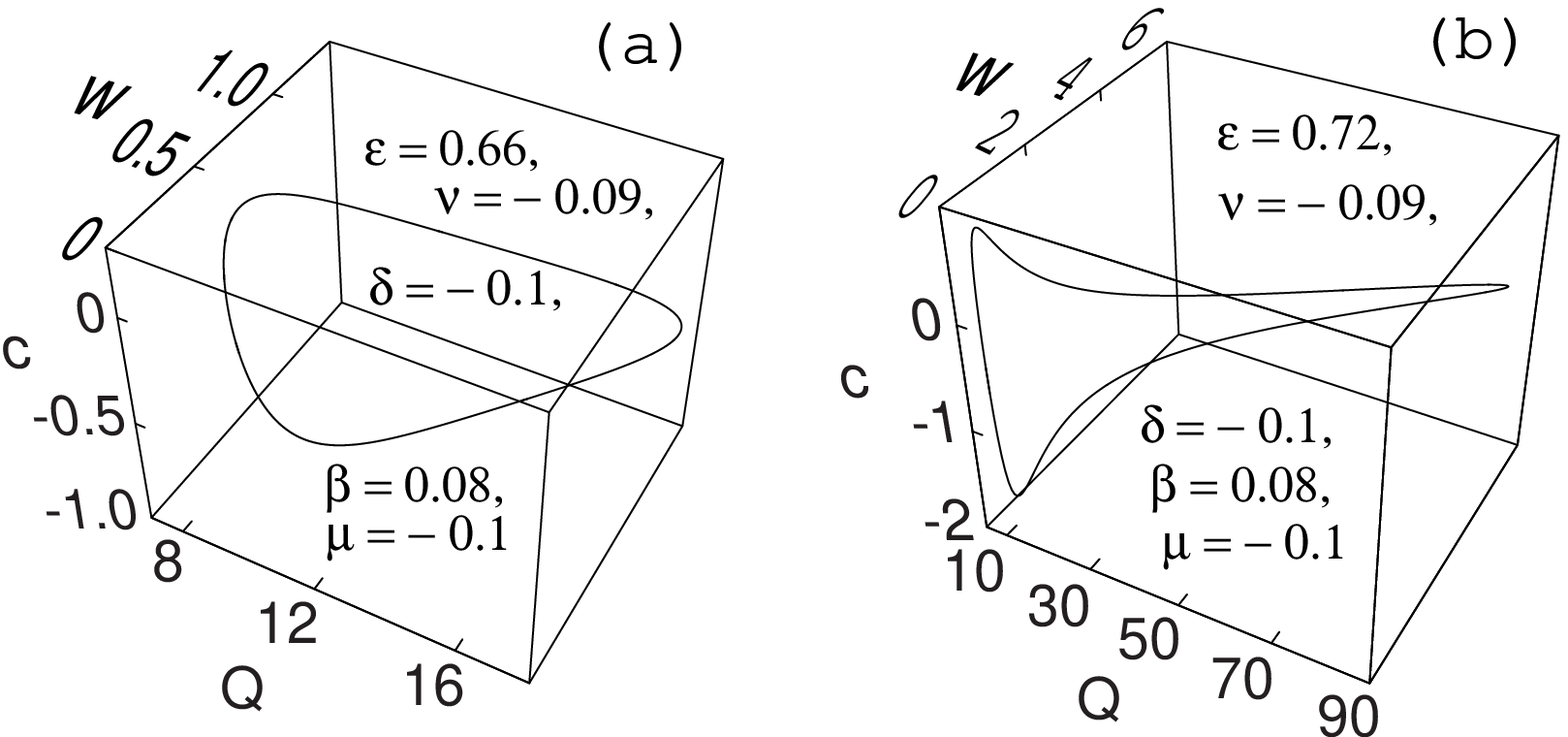}}
\caption{Two examples of stable limit cycle of the system (\ref{Model}) in
($Q, w, c$)-space. The parameters are shown in (a) and (b)
respectively. Note the different scales for the axes in (a) and (b).}
\label{Fig:attr1}
\end{figure}

Curve $1$ in Fig.\ref{Fig:ne1}a is the bifurcation boundary
(threshold) where the stable fixed point turns into an unstable one.
Above curve $1$, $\mbox{Re}[\lambda_{1,2}] >0$,
$\mbox{Im}[\lambda_{1}] = -\mbox{Im}[\lambda_{2}]$,
$\mbox{Re}[\lambda_{3}] <0$ and $\mbox{Im}[\lambda_{3}] =0$.
On line $1$, a stable FP  is transformed into a stable limit
cycle (LC) and an unstable FP. We confirmed, numerically, that the
{\em stable} limit cycle of model~(\ref{Model}) does indeed exist
between the solid curves $1$ and $2$.  Two examples of LC in
($Q, w, c$)-space are shown in Fig.~\ref{Fig:attr1}a and \ref{Fig:attr1}b.
The limit cycle has zero size on line $1$ in Fig.\ref{Fig:ne1}a. This corresponds
to a super-critcal Hopf bifurcation \cite{Guckenheimer}.
The limit cycle elongates in $Q$ axis direction when $\epsilon$
increases. It becomes infinitely large and breaks off on line $2$ in
Fig.\ref{Fig:ne1}a.

The smaller region in Fig.\ref{Fig:ne1}a surrounded by the light solid curve, is
copied from Fig.\ref{Fig:ne1}b  for comparison. This is the
boundary for the existence of pulsating solitons found by direct numerical
simulations of CGLE (\ref{CGLE}). One can see that the thresholds (solid
curves $1$ and $2$) of the existence of the stable LC provide reasonably good
estimates for the boundaries.

\begin{figure}[c]
\centering
\epsfxsize=6cm \mbox{\epsffile{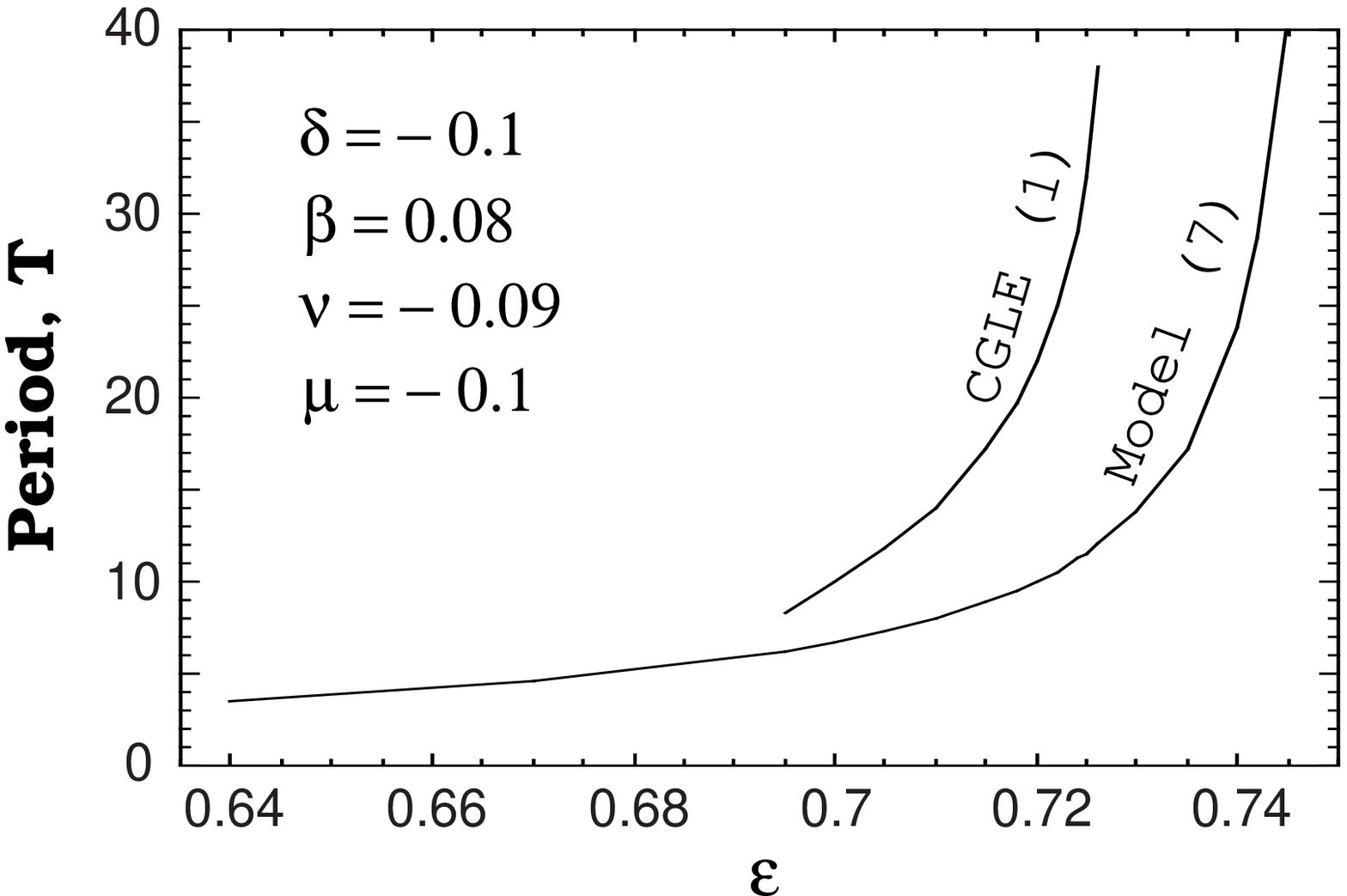}}
\caption{Period of pulsations $T$ versus $\epsilon$.
Upper curve correspond to exact numerical results while the lower
curve describes the data obtained from the low-dimensional model.}
\label{Fig:front}
\end{figure}

The period of oscillations, $T$, depends on the parameters of the system.
A plot of $T$ versus $\epsilon$ at constant
$\nu= - 0.09$ is given in Fig.\ref{Fig:front}.
The upper curve represents the results of exact numerical simulations
of the CGLE, while the lower curve shows the period of oscillations
found from our finite-dimensional model.
There is an apparent difference in the values found for the period,
due to the drastic reduction in the number of degrees of freedom in the model.
However, the two curves have the same qualitative behaviour.
In particular, each curve starts with a finite value of the period $T$
at the lower boundary of the region where pulsating solitons
exist, and increases to infinity at the upper boundary.

Above the curve $2$,
the soliton amplitude $A= [Q/(2w)]^{1/2}$ remains almost
constant, but the soliton energy $Q$ and the width $w$ increase
monotonically with $t$. This motion is tantamount to the localized
solution of CGLE with constant amplitude and the width
that increases indefinitely. The final stage of this motion is an
asymptotic transformation of the soliton into two separating fronts
with constant velocities.
Thus, the qualitative agreement above the line $2$ is also
fairly good and the main features of the
dynamics are captured correctly.

\begin{figure}[c]
\centering
\epsfxsize=8.5cm \mbox{\epsffile{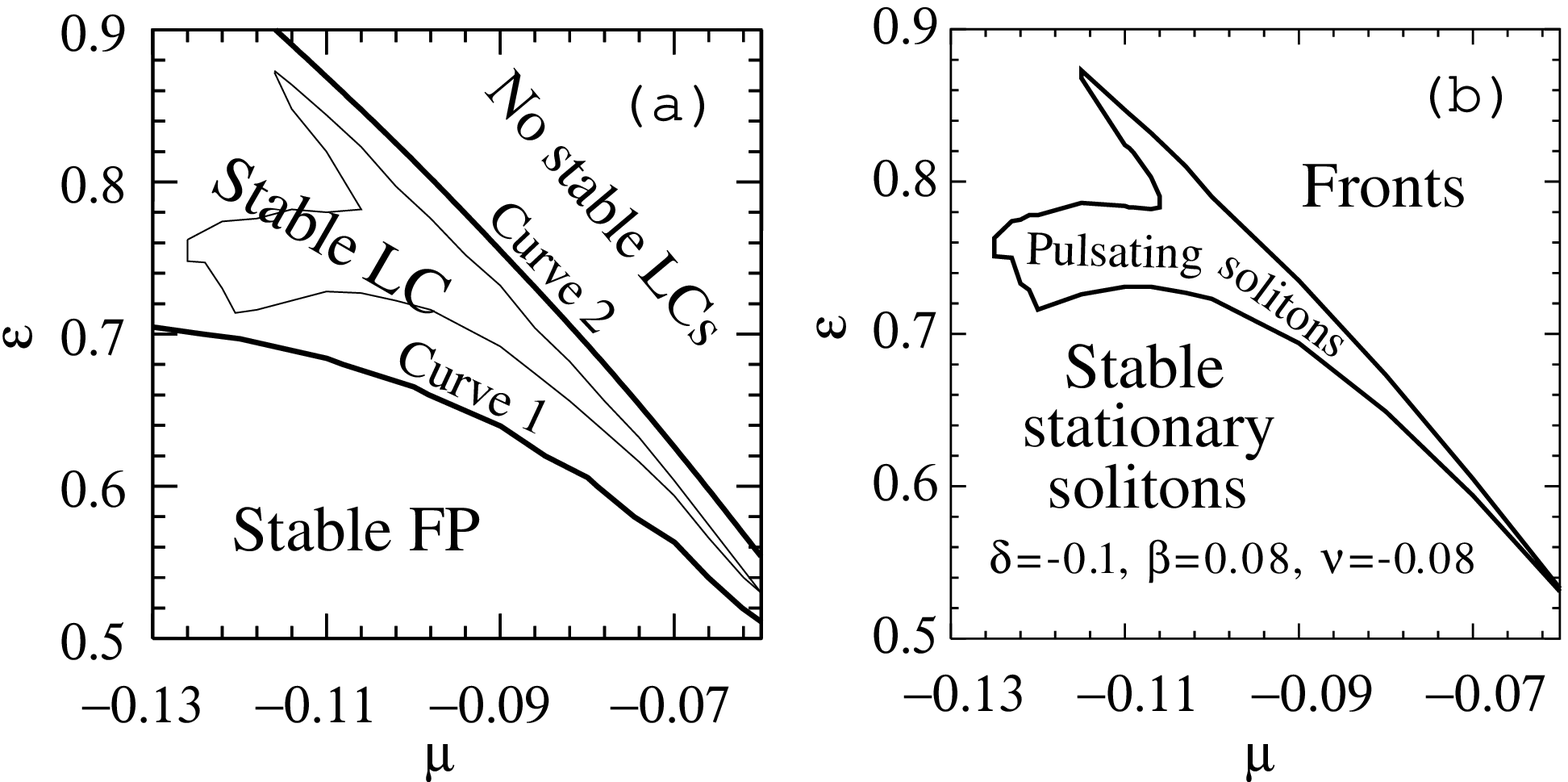}}
\caption{
(a) Diagram of existence and stability of FPs and LCs of the
reduced model in the ($\mu, \epsilon$)-plane.  Stable LCs
exist between the curves $1$ and $2$.
(b) Existence diagram based on numerical simulations of CGLE
\protect \cite{AST01}. The solid curve bounding the region for
pulsating solitons in (b) is copied to (a) for comparison.
The parameters of the dynamical system are shown in (b).}
\label{Fig:me1}\end{figure}

Another slice of the parameter space, namely the ($\mu, \epsilon$)-plane,
is shown in Fig.~\ref{Fig:me1}a. As in Fig.~\ref{Fig:ne1}a, there are
two FPs  in the region presented. One of them is always unstable. The
stable FP exists below the solid curve $1$. The bifurcation on the
curve $1$ is the same as that in Fig.~\ref{Fig:ne1}a. The solid curve
$2$ corresponds to the transition between limit cycles and the
solutions with increasing width. From Fig.~\ref{Fig:ne1} and
Fig.~\ref{Fig:me1}, one can see that the results obtained from the
model~(\ref{Model}) for pulsating solitons and the boundaries for their
existence are in reasonable agreement with direct numerical simulations
of the CGLE over a wide range of the system parameters.

Thus, considering the evolution of the dissipative soliton profile, the
model~(\ref{Model}) is able to predict: (a) stable and unstable
stationary solitons; (b) periodic soliton pulsations; (c) the unlimited
increase in the soliton width at constant amplitude (which is equivalent to
splitting solitons into moving fronts); and (d) bifurcations between
these dynamical behaviours.

In summary, we have derived a finite-dimensional dynamical system
associated with the CGLE. We have demonstrated that the pulsating
solitons of the CGLE correspond to the limit cycles of this system,
while stationary solitons of the CGLE are related to the
fixed points. We have found the approximate bifurcation boundaries
between different types of solutions in the parameter space.

\begin{acknowledgments}
This work was funded by the Australian Research Council. The authors
are grateful to Dr. A. Ankiewicz for a critical reading of the manuscript.
\end{acknowledgments}

\end{document}